\documentclass[aps,pra]{revtex4}

\usepackage{graphicx}
\usepackage{amsmath}
\usepackage{amssymb}
\usepackage{dcolumn}
\usepackage{bm}
\usepackage{multirow}
\usepackage{enumitem}
\usepackage{booktabs}
\usepackage{xcolor}

\begin{document}

\title{Strong Local Passivity in Unconventional Scenarios:\\ A New Protocol for Amplified Quantum Energy Teleportation}

\author{Songbo Xie}
\affiliation{Department of Electrical and Computer Engineering, North Carolina State University, Raleigh, North Carolina 27606, USA }

\author{Manas Sajjan}
\affiliation{Department of Electrical and Computer Engineering, North Carolina State University, Raleigh, North Carolina 27606, USA }

\author{Sabre Kais}
\email{skais@ncsu.edu}
\affiliation{Department of Electrical and Computer Engineering, North Carolina State University, Raleigh, North Carolina 27606, USA }

\date{\today}

\begin{abstract}
Quantum energy teleportation (QET) has been proposed to overcome the restrictions of strong local passivity (SLP) and to facilitate energy transfer in quantum systems. Traditionally, QET has only been considered under strict constraints, including the requirements that the initial state be the ground state of an interacting Hamiltonian, that Alice’s measurement commute with the interaction terms, and that entanglement be present. These constraints have significantly limited the broader applicability of QET protocols. In this work, we demonstrate that SLP can arise beyond these conventional constraints, establishing the necessity of QET in a wider range of scenarios for local energy extraction. This leads to a more flexible and generalized framework for QET. Furthermore, we introduce the concept of a ``local effective Hamiltonian,'' which eliminates the need for optimization techniques in determining Bob's optimal energy extraction in QET protocols. As an additional advantage, the amount of energy that can be extracted using our new protocol is amplified to be 7.2 times higher than that of the original protocol. These advancements enhance our understanding of QET and extend its broader applications to quantum technologies. To support our findings, we implement the protocol on quantum hardware, confirming its theoretical validity and experimental feasibility.
\end{abstract}
\maketitle

\section{Introduction}
Quantum state teleportation (QST) is widely recognized for its ability to transfer quantum states to distant locations~\cite{bennett1993teleporting,bouwmeester1997experimental,furusawa1998unconditional}. However, it is well understood that the energy associated with a quantum state cannot be teleported, as~the sender transmits only information about the state. Using this information, along with classical communication and shared entanglement, the~receiver reconstructs the quantum state. The~energy required for this reconstruction is supplied locally by the receiver, rather than being teleported from the~sender.

In contrast, a~related protocol, quantum energy teleportation (QET), was believed to enable the teleportation of energy~\cite{hotta2009quantum,hotta3954quantum}. Similarly to QST, QET relies on classical communication. However, QET protocols impose stricter constraints compared to QST. While QST only requires the sender and receiver to share an entangled state, QET further requires that this state be the ground state of an interacting Hamiltonian. Additionally, in~QST, the~sender performs a measurement and transmits the outcome to the receiver via a classical communication channel. In~QET, however, the~observable used for the measurement must also commute with the interaction term in the Hamiltonian, ensuring that the local energy of the receiver remains undisturbed prior to the teleportation~process.

Since its initial discovery in spin-chain systems~\cite{hotta2008protocol,hotta2009quantum}, QET has been extensively explored in many other physical systems. Theoretical studies have investigated its application in relativistic quantum field theory~\cite{hotta2008quantum,ikeda2023criticality}, trapped ions~\cite{hotta2009quantumtrapped}, harmonic oscillators~\cite{hotta2010quantum}, black-hole physics~\cite{hotta2010controlled}, linear harmonic chains~\cite{nambu2010quantum}, quantum Hall systems~\cite{yusa2011quantum}, Gibbs spin particles~\cite{frey2013quantum,trevison2015quantum}, squeezed vacuum states~\cite{hotta2014quantum}, topological orders~\cite{ikeda2023investigating}, and~quantum information science~\cite{ikeda2024quantum}. More recently, experimental efforts have demonstrated progress toward realizing QET, including in an NMR system~\cite{rodriguez2023experimental} and on superconducting quantum hardware~\cite{ikeda2023demonstration}. The~combination of QST and QET was proposed to enable a long-range QET protocol~\cite{ikeda2023long}. A~trade-off relationship between QST and QET was proposed, suggesting their potential exclusivity~\cite{wang2024quantum}. In~\cite{fan2024strong}, a~relation between QET and quantum steering was~suggested.

However, despite its name, the~QET protocol does not physically teleport energy between distant locations. Instead, the~sender's classical message only enables the receiver to access energy that was previously inaccessible prior to the implementation of the protocol~\cite{rodriguez2023experimental}. This argument can be further supported in~\cite{wu2024strong}, where the receiver’s extracted energy can exceed the sender’s injected energy, indicating that the extracted energy cannot originate from the sender. This inaccessibility of energy extraction can be rigorously described by the concept of \textit{strong local passive} 
 (SLP) states~\cite{frey2014strong,alhambra2019fundamental}. An~SLP state is a multipartite quantum state, denoted by $\rho$, where no local quantum operation $\mathcal{G}_B$ applied to a subsystem can extract energy from the total system, characterized by the Hamiltonian $H$. Mathematically, this is expressed as the nonnegativity of the energy difference:
\begin{equation}\label{deltae}
    \Delta E = \text{Tr}[H(\mathcal{I}_A \otimes \mathcal{G}_B)\rho] - \text{Tr}(H\rho) \geq 0,\quad \forall\mathcal{G}_B.
\end{equation}
In a QET protocol, the~two parties share an SLP state, preventing the receiver from locally extracting energy. The~sender then measures its local qubit, but~the two-qubit system remains in an SLP state, meaning the receiver still cannot extract energy. However, classical communication between the sender and receiver breaks this limitation, making $\Delta E$ negative. A~negative $\Delta E$ indicates that the receiver can locally extract energy, overcoming the restriction imposed by~SLP.

However, previous QET protocols have been studied only under scenarios with strict constraints on the initial shared state and/or on the sender’s measurement observables. In~our view, these constraints arise from the initial misconception---reinforced by the name ``quantum energy teleportation''---that QET directly teleports energy between distant locations, as~well as from insufficient attention to the connection between QET and~SLP.

In this work, we demonstrate that SLP can arise beyond these conventional constraints, establishing the necessity of QET in a wider range of scenarios for local energy extraction. Specifically, by~introducing the concept of a ``local effective Hamiltonian,'' we construct an SLP state that satisfies none of these restrictions, yet still forbids local energy extraction. As~a result, energy extraction is only possible through a QET protocol. In~this sense, relaxing these constraints reveals that situations where energy remains locally inaccessible are more common than previously thought, broadening the applicability of QET. As~a separate and technical advantage, the~local effective Hamiltonian simplifies the identification of optimal operations for the receiver, enabling maximal energy extraction without the need for cumbersome optimization techniques. Therefore, our results not only deepen the understanding of QET and its connection to SLP states but also expand its potential applications in quantum science and~technology.

Finally, we implement our new protocol on quantum hardware, confirming both its theoretical correctness and experimental~feasibility.

\section{Quantum Energy Teleportation and Locally Inaccessible~Energy}
To begin, we revisit the minimal model of QET introduced in~\cite{hotta2010energy}, 
which involves two qubits, $A$ and $B$, shared between the sender (Alice) and the receiver (Bob). The~Hamiltonian of the system is given by
\begin{equation}\label{hab}
    H_{AB} = -hZ_A - hZ_B + 2\kappa X_A \otimes X_B,
\end{equation}
where $h$ and $\kappa$ are positive constants, and~$X_i$ and $Z_i$ (with $i \in \{A,B\}$) are the Pauli-$X$ and Pauli-$Z$ operators for qubit $i$. The~two qubits are initially prepared in the ground state $|g\rangle$ of $H_{AB}$, expressed as
\begin{equation}\label{g1}
    |g\rangle = \cos(\theta)|00\rangle_{AB} - \sin(\theta)|11\rangle_{AB},
\end{equation}
where $\tan(2\theta) \equiv \kappa/h$. Since the ground state has the lowest possible energy, neither Alice nor Bob can extract any energy from $|g\rangle$.  

To enable energy extraction, the~QET protocol requires Alice to measure her local qubit using the operators $\{|+\rangle\langle+|, |-\rangle\langle-|\}$, where $|\pm\rangle = (|0\rangle \pm |1\rangle)/\sqrt{2}$. Alice's measurement pushes the system away from the ground state $|g\rangle$, thus injecting energy into the system. But~the measurement operators commute with the interaction Hamiltonian:  
\[
[\ |+\rangle\langle+|, 2\kappa X_A\otimes X_B\ ] = [\ |-\rangle\langle-|, 2\kappa X_A\otimes X_B\ ] = 0.
\]  
This commutativity ensures that Alice's injected energy stays localized with her. Specifically, the~measurement only affects the expectation value of the $-hZ_A$ term in Equation~\eqref{hab}, which corresponds to Alice's qubit. In~contrast, the~terms associated with Bob, $-hZ_B + 2\kappa X_A \otimes X_B$, remain~unaffected.


Notably, after~Alice's measurement, the~state of the system becomes
\begin{equation}\label{slp}
    \rho_\text{SLP} = \frac{1}{2} \Big( |+\rangle\langle+|_A \otimes |b^+\rangle\langle b^+|_B \;+\; |-\rangle\langle-|_A \otimes |b^-\rangle\langle b^-|_B \Big),
\end{equation}
with $|b^\pm\rangle = \cos(\theta)|0\rangle \mp \sin(\theta)|1\rangle$ representing Bob's local states after the measurement. The~state $\rho_\text{SLP}$ is an SLP state, meaning that Bob cannot extract energy from it using any local general operation $\mathcal{G}$, as~explained~earlier.  

For Bob to access the inaccessible energy within Equation~\eqref{slp}, the~QET protocol requires Alice to communicate her measurement outcome to Bob, based on which Bob applies a conditional operator,
\begin{equation}  
G_\pm \equiv \exp[\pm i(\phi - \theta)Y],  
\end{equation}  
to his qubit, where $\tan(2\phi) \equiv 2\kappa/h$, with~the sign $\pm$ depending on Alice's measurement outcome. $Y$ is the Pauli-Y operator. The~resulting state, expressed as
\begin{equation}
    \rho_f = \frac{1}{2} \left[ |+\rangle\langle+|_A \otimes \left(G_+|b^+\rangle\langle b^+|_BG_+^\dagger\right) \;+\; |-\rangle\langle-|_A \otimes \left(G_-|b^-\rangle\langle b^-|_BG_-^\dagger\right) \right],
\end{equation}
exhibits a negative energy difference:
\begin{equation}\label{extractold}
\Delta E = \text{Tr}(H_{AB} \rho_f) - \text{Tr}(H_{AB} \rho_\text{SLP}) = -2\sin^2(\phi - \theta)\sqrt{h^2+4\kappa^2} < 0.  
\end{equation}  
This negativity indicates that Bob’s conditional operation successfully breaks the limit of strong local passivity, allowing him to extract the previously inaccessible energy that would remain unavailable without Alice’s classical communication. For~our parameter setting, $h=1$ and $\kappa=1.5$, which will be used in the next section, the~extracted energy for this protocol is~0.1114.

It is important to clarify that after Alice's measurement, the~SLP property of the system is fragile and can vanish as the system evolves under \( H_{AB} \). Consequently, Bob's local energy extraction is, in~principle, possible even without Alice's message, provided he waits for an appropriate duration for the system to evolve and then applies a final local operation to extract energy. Nevertheless, we emphasize that the key advantage of QET lies in its speed. When relativistic constraints are not considered, Alice's message can travel much faster than the natural timescale of the system's evolution \cite{diener1997energy}. This enables Bob to extract energy almost instantaneously, distinguishing QET from conventional energy extraction~mechanisms.

We point out that the following constraints have been widely assumed for previous QET protocols. However, these constraints are intrinsically irrelevant to the SLP property of quantum states and should not be assumed as fundamental for applications of QET. Yet, they still appear consistently in discussions of QET. In~this work, we demonstrate that these constraints are not required for a state to exhibit SLP. Consequently, even when these constraints are relaxed, QET remains essential for enabling local energy extraction. The~constraints are as follows:

\begin{enumerate}[label=(\alph*)]
    \item \textbf{Ground State}:  
It has been argued that if a density matrix $\rho$ commutes with the Hamiltonian, $[\rho, H_{AB}] = 0$, and~the population of the ground state $|g\rangle$ exceeds a threshold $p^*$ (as defined in~\cite{frey2014strong}), then $\rho$ is an SLP state. Consequently, most existing QET protocols initialize the system in the ground state $|g\rangle$ of $H_{AB}$, ensuring that the state after Alice's measurement retains sufficient population in $|g\rangle$ to preserve the SLP property. However, this condition is only sufficient, not necessary. We show that a system can exhibit SLP even when initialized in an excited pure state with zero population in $|g\rangle$. Moreover, after~Alice's measurement, the~resulting state can still be an SLP state, requiring Bob to rely on Alice’s information for energy extraction. While it is intuitive that energy extraction is forbidden from a ground state, it is counter-intuitive that energy can also be blocked for certain excited states. This suggests that the common understanding that only ground states prevent energy extraction may be~incomplete.

    \item \textbf{Commutativity}: To construct an SLP scenario that prevents Bob from locally extracting energy, it is typically assumed that Alice’s measurement does not increase the energy in Bob’s surroundings, which further requires Alice’s measurement observable to commute with the interaction term in \(H_{AB}\), ensuring that her measurement does not inject energy into Bob’s surroundings. However, we emphasize that the SLP property stands on its own and is independent of this assumption. We demonstrate that even when Alice’s measurement observable does not commute with the interaction term, the~resulting state can still exhibit SLP. Consequently, Bob still requires the QET protocol for local energy~extraction.

    \item \textbf{Entanglement}: It was believed that the shared state between Alice and Bob must be entangled for Alice's measurement to have any influence on Bob's subsystem. It was believed that {``\it a large amount of teleported energy requests a large amount of consumption of the ground-state entanglement between A and B in this model''} \cite{hotta3954quantum}. Relations were further studied between the entanglement breaking due to Alice's measurement and the teleported energy~\cite{hotta2010energy}. Arguments also suggested that if the ground state is entangled and Alice's measurement disentangles the system, it becomes impossible for Bob, using only local operations and classical communications, to~restore the entangled ground state and extract energy~\cite{alhambra2019fundamental}. Along the same line, it was also suggested that quantum resources can improve the energy extraction efficiency for QET~\cite{fan2024role}. However, we demonstrate that entanglement is not a strict requirement for the exhibition of SLP and the application of QET. In~fact, in~our proposed protocol, the~shared state between Alice and Bob remains disentangled throughout the entire process, yet the system still exhibits the SLP property. To~understand this, we introduce the concept of a \textit{local effective Hamiltonian}. We show that Alice's actions can influence Bob’s subsystem not only by changing the ``departure state" through entanglement but also by changing the ``destination state" via Bob's local effective Hamiltonian. As~long as either the departure or destination state is affected, conditional operations are required to break SLP for Bob's local energy extraction. Therefore, the~necessity of entanglement can be relaxed if Alice’s measurements can influence the destination state alone.
\end{enumerate}

We note that some previous works have gone beyond one or two of the constraints discussed above. For~example, Ref.~\cite{frey2013quantum} considered a QET protocol with an initial Gibbs state that may contain no entanglement, while Ref.~\cite{wu2024strong} introduced a class of initial states that are not ground states. However, in~both cases, the~SLP condition was analyzed only before Alice's measurement---not afterward. It can be readily verified that their post-measurement states can violate the SLP condition, meaning that classical communication is not required for Bob’s energy extraction. As~a result, these protocols do not truly require the implementation of QET. In~contrast, our work not only emphasizes the importance of verifying the SLP condition after Alice’s measurement but also relaxes all three constraints within one single and minimal~protocol.

\section{Strong Local Passivity Relaxing Previous~Constraints}

We propose a new QET protocol involving two qubits, $A$ and $B$, shared between Alice and Bob, under~the following flip-flop Hamiltonian:
\begin{equation}\label{flipflop}
\begin{split}
    H_{AB} &= -hZ_A - hZ_B + 2\kappa(\sigma_+ \otimes \sigma_- + \sigma_- \otimes \sigma_+) \\
           &\equiv -hZ_A - hZ_B + \kappa(X_A \otimes X_B + Y_A \otimes Y_B),
\end{split}
\end{equation}  
where $Z$, $X$, and~$Y$ represent the Pauli matrices, and~$\sigma_\pm\equiv(X\mp iY)/2$ are the qubit raising and lowering~operators.  

Compared to Equation~\eqref{hab}, this Hamiltonian replaces one of the two $X_A\otimes X_B$ terms with a $Y_A \otimes Y_B$ interaction term. Consequently, the~total interaction term, $(X_A \otimes X_B + Y_A \otimes Y_B)$, cannot be factorized. As~we will show later, this replacement will greatly enhance the amount of energy that can be extracted by~Bob.

We consider the deep strong coupling regime~\cite{forn2019ultrastrong}, setting $\kappa > h > 0$. In~this regime, the~Hamiltonian in Equation~\eqref{flipflop} has four eigenstates with corresponding eigenvalues:
\begin{equation}\label{4eigenstates}
    \begin{split}
        (|01\rangle - |10\rangle)/\sqrt{2}, &\quad -2\kappa, \\
        |00\rangle, &\quad -2h, \\
        |11\rangle, &\quad +2h, \\
        (|01\rangle + |10\rangle)/\sqrt{2}, &\quad +2\kappa.
    \end{split}
\end{equation}
We choose the first excited state, $|00\rangle$, as~the initial state. Alice then measures her qubit $A$ using the operators $\{|+\rangle\langle+|, |-\rangle\langle-|\}$. Regardless of the outcome, Bob's qubit remains in state $|0\rangle$, as~the qubits are disentangled. The~density matrix after Alice's measurement is
\begin{equation}\label{slp2}
    \rho_\text{SLP} = \frac{1}{2}\left(|0\rangle \langle 0|_A\otimes|0\rangle\langle0|_B + |1\rangle \langle 1|_A\otimes|0\rangle\langle0|_B\right).
\end{equation}
To verify whether this state is SLP for Bob according to the Hamiltonian Equation~\eqref{flipflop}, one would need to check the inequality in Equation~\eqref{deltae}, which involves parameterizing Bob's general operation with 12 parameters for a qubit completely positive trace-preserving (CPTP) map~\cite{nielsen2010quantum,boyd2004convex,watrous2018theory, sim2020user}. However, a~necessary and sufficient condition for determining whether a state $\rho$ is SLP for a Hamiltonian $H$ is given by a $4 \times 4$ matrix $M(\rho, H)$, the~form of which is defined in~\cite{alhambra2019fundamental} and given in Appendix \ref{appa}. Specifically, $\rho$ is SLP for $H$ if and only if $M(\rho, H)$ is positive~semi-definite. 

In our case, it is straightforward to verify that both $M(|00\rangle\langle00|, H_{AB})$ and $M(\rho_\text{SLP}, H_{AB})$ share the identical eigenvalue set, $\{2, 0, 0, 0\}$, all of which are nonnegative, confirming that these two matrices are positive semi-definite. Thus, starting from the first excited state $|00\rangle$, Bob cannot extract energy locally. After~Alice's measurement, the~state evolves to Equation~\eqref{slp2}, and~Bob still cannot extract energy locally. Instead, Bob needs communication from Alice for his energy~extraction.

In contrast, if~we had started with any of the other three eigenstates in Equation~\eqref{4eigenstates}, including the ground state $(|01\rangle-|10\rangle)/\sqrt{2}$, the~resulting states after Alice's measurement would not maintain the SLP property. Specifically, starting from the ground state, and~performing Alice's measurement, we compute the $M$ matrix of the resulting state and find its four eigenvalues to be $\{2, 1.5,-0.5,0\}$. This confirms that the post-measurement state is not an SLP state, meaning QET protocols are not necessary for Bob's energy extraction. Therefore, our initial choice of the first excited state $|00\rangle$ is well~justified.

Through the new protocol, we observe that all three previously considered constraints for QET are relaxed. First, the~system is initialized in an excited state where the ground-state population is zero. Second, although~Alice's measurement does not commute with the interaction term $Y_A\otimes Y_B$, it does not inject energy into this term, as~$\langle Y_A\otimes Y_B\rangle$ remains 0 throughout the entire process. Finally, the~system begins in the product state $|00\rangle$, meaning Alice's measurement cannot directly affect Bob's local~state. 

At first glance, it may seem puzzling that Alice's measurement is needed in a product-state construction, as~the measurement cannot affect Bob's local state. To~address this, below we introduce the concept of a local effective Hamiltonian for Bob. We show that, for~energy extraction, Bob must rotate his local state to the ground state of this effective Hamiltonian. In~a product-state setup, Alice’s measurement does not alter Bob’s initial (“departure”) state, but~it does modify the effective Hamiltonian---and thus the ground (“destination”) state of this Hamiltonian that Bob must reach. As~a result, classical communication from Alice remains necessary to inform Bob of the appropriate rotation. In~general, conditional operations are required to break the SLP constraint as long as one of the departure or destination states is contingent on Alice's measurement. Notably, this allows the relaxation of the entanglement requirement when Alice’s measurement only influences the destination~state.

Now we introduce Bob's local effective Hamiltonian. Since the two qubits remain disentangled after Alice's measurement, the~total energy of the system is given by
\begin{equation}
    \langle H_{AB}\rangle = -h\langle Z_A\rangle_A - h\langle Z_B\rangle_B + \kappa\langle X_A\rangle_A\langle X_B\rangle_B + \kappa\langle Y_A\rangle_A\langle Y_B\rangle_B,  
\end{equation}  
where $\langle \cdot \rangle_A \equiv \langle \psi_A | \cdot | \psi_A \rangle$ denotes the expectation value of the operator $\cdot$ with respect to Alice's state $|\psi_A\rangle$, and~similarly for $\langle \cdot \rangle_B$ with respect to Bob's~state.

If Alice’s measurement yields the outcome $+$, her state collapses to $|+\rangle$, and~the total energy becomes
\begin{equation}
    \langle H_{AB} \rangle = -h\langle Z_B \rangle_B + \kappa\langle X_B \rangle_B,  
\end{equation}  
implying that Bob's effective Hamiltonian is $-hZ_B + \kappa X_B$. Similarly, if~Alice's outcome is $-$, Bob's effective Hamiltonian becomes $-hZ_B - \kappa X_B$.

One might wonder whether this effective Hamiltonian has a dynamical effect on Bob's qubit. The~answer is yes. To~show this, Alice fixes her state after the measurement using the quantum Zeno effect~\cite{misra1977zeno,itano1990quantum,fischer2001observation}, by~repeatedly measuring her state at a rate much faster than the timescale of $H_{AB}$, which is commonly characterized by $\max(h,\kappa)$.

By assuming that Alice's measurement outcome is $+$ (the analysis for $-$ follows similarly), the~general two-qubit state is given by
\begin{equation}  
    |\psi(t)\rangle = |+\rangle_A \otimes |\varphi(t)\rangle_B.
\end{equation}  
Under Schr\"odinger's equation, the~time evolution of the state is
\begin{equation}  
    |\psi(t+dt)\rangle = \Big(|+\rangle\langle+| \otimes \mathbb{I}\Big) \Big(\mathbb{I} - iH_{AB}dt\Big) |\psi(t)\rangle,  
\end{equation}  
where the second term represents free evolution under \( H_{AB} \), and~the first term enforces the projective operator due to Alice's repeated~measurements.

Using $\langle+|X|+\rangle = 1$ and $\langle+|Y|+\rangle = \langle+|Z|+\rangle = 0$, the~evolution simplifies to
\begin{equation}  
    |\varphi(t+dt)\rangle = \Big(\mathbb{I} - i(-hZ_B + \kappa X_B)dt\Big) |\varphi(t)\rangle,  
\end{equation}  
or, equivalently,
\begin{equation}  
    i\dfrac{\partial}{\partial t}|\varphi(t)\rangle = \Big(-hZ_B + \kappa X_B\Big) |\varphi(t)\rangle.  
\end{equation}  

Similarly, for~the $-$ outcome, Bob's effective Hamiltonian becomes $-hZ_B - \kappa X_B$. Thus, the~local effective Hamiltonian for Bob is
\begin{equation}\label{effective}
    H_B^{(\pm)} = -hZ_B \pm \kappa X_B.  
\end{equation}

For Bob to extract energy using Alice's information, we diagonalize the effective Hamiltonian Equation~\eqref{effective}. The~ground states $|g^\pm\rangle$ and excited states $|e^\pm\rangle$ of the effective Hamiltonian, together with their eigenvalues, are given by
\begin{equation}
\begin{split}
    -\sqrt{h^2 + \kappa^2}, \quad |g^\pm\rangle \equiv&\ \ \ \cos(\theta)|0\rangle \mp \sin(\theta)|1\rangle, \\
    +\sqrt{h^2+\kappa^2},\quad |e^\pm\rangle\equiv&\pm\sin(\theta)|0\rangle + \cos(\theta)|1\rangle.
\end{split}
\end{equation}  
where $\tan(2\theta) = \kappa/h$. The~effective Hamiltonian can be reformulated as
\begin{equation}
    H_B^{(\pm)}=\sqrt{h^2+\kappa^2}\Big(|e^\pm\rangle\langle e^\pm|-|g^\pm\rangle\langle g^\pm|\Big).
\end{equation}

We now highlight an additional advantage of introducing the local effective Hamiltonian \eqref{effective}. In~general QET protocols, determining Bob's optimal conditional operations for maximal energy extraction required cumbersome optimization techniques, typically involving the parameterization of a general single-qubit unitary operation (for example, see~\cite{haque2024aspects}). However, once the conditional Hamiltonian is established, Bob's optimal operations simply correspond to unitary transformations that rotate his current state to the conditional ground states $|g^\pm\rangle$ of the effective~Hamiltonians.

Specifically, in~our case, to~extract energy, Bob applies to his qubit the conditional RY gate, \( RY(\mp 2\theta) = \exp\left(\pm 2i\theta Y\right) \), which rotates his initial state $|0\rangle\equiv \cos(\theta)\pm\sin(\theta)|1\rangle$ to the corresponding ground state $|g^\pm\rangle$.  This highlights the role of the effective Hamiltonian: although Alice and Bob are initially unentangled, Alice's measurement influences Bob by modifying his destination state via the effective Hamiltonian. Thus, entanglement is not required in this QET protocol. As~can be verified, the~amount of energy that Bob extracts is
\begin{equation}\label{extract}
    E_\text{extract}=2\sin^2(\theta)\sqrt{h^2+\kappa^2}.
\end{equation}

\begin{figure}[t]
    \includegraphics[width=0.4\linewidth]{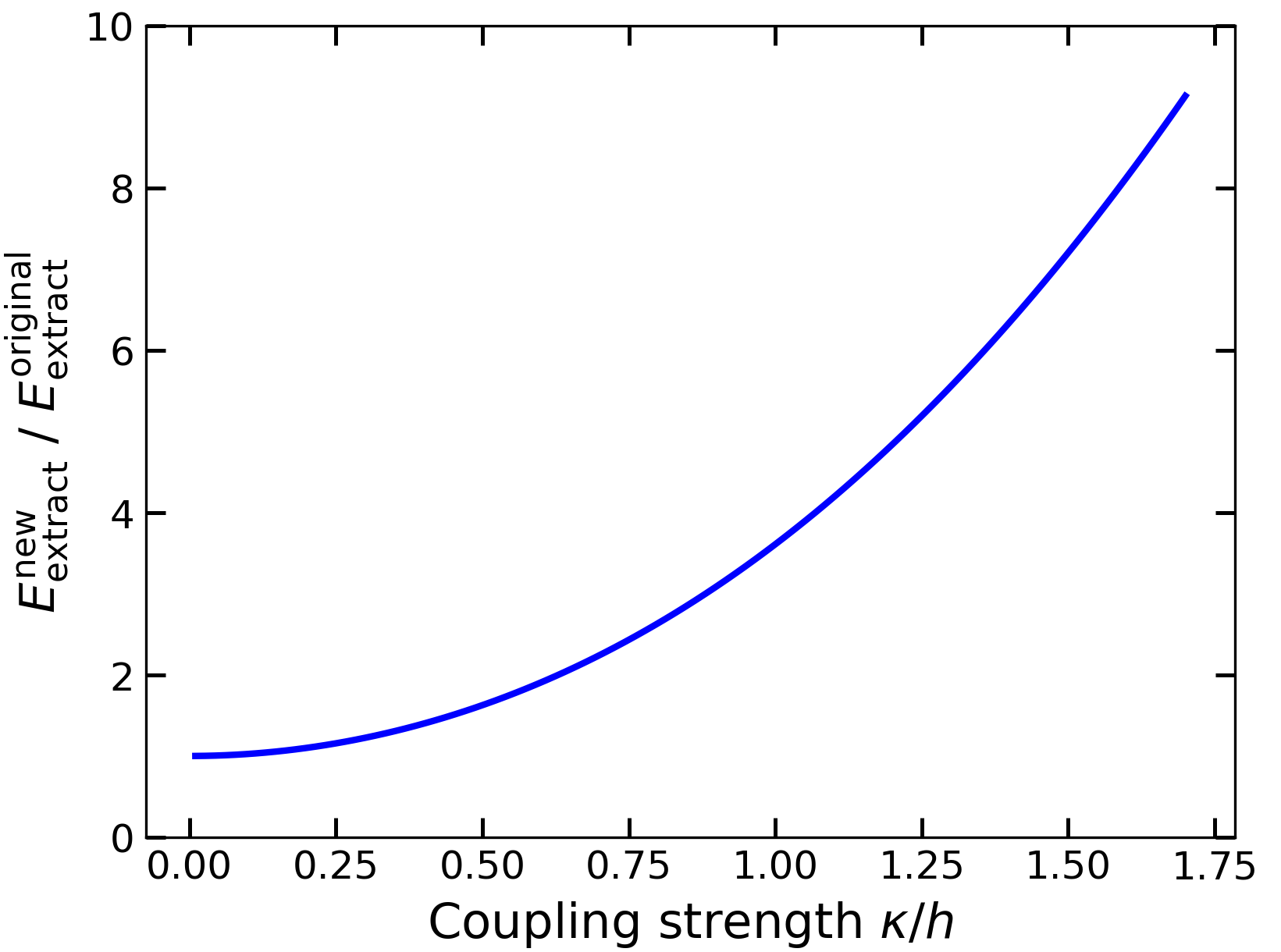}
    \caption{The ratio of Bob's extracted energy for the new protocol, $E_{\mathrm{extract}}^{\mathrm{new}}$,  and~Bob's extracted energy for the original protocol, $E_{\mathrm{extract}}^{\mathrm{original}}$, as~a function of the coupling strength $\kappa$ in {units} 
    	of $h$.}
    \label{fig:ratio}
\end{figure}

For our parameter setting, $h=1$ and $\kappa=1.5$, which will be used in the next section, the~extracted energy is amplified, reaching up to 0.8028, which is 7.2 times higher than that of the original QET protocol, where the extracted energy was only 0.1114. This further highlights the advantage of our newly proposed QET~protocol.

One might argue that the observed enhancement in energy extraction simply results from the use of an excited initial state in the new protocol, in~contrast to the use of the ground state in the original protocol, which naturally provides more available energy for Bob to extract. This raises an important question: under what standard should we understand the enhancement in energy extraction?

We emphasize that the use of the initial state in the new protocol depends on the choice of system parameters. Our intention in selecting the excited state was simply to demonstrate that such a state can also exhibit the SLP property. However, in~general, one can tune the parameters so that $\kappa < h$, making the same state $|00\rangle$ the ground state of the system. In~this case, the~SLP property still holds: the state $|00\rangle$ remains SLP both before and after Alice’s measurement in the Pauli-$X$ basis. As~a result, Bob still requires the QET protocol to extract local energy, and~all subsequent discussions remain valid. In~this alternative scenario, we can directly compare the amount of energy Bob can extract in the two protocols, as~shown in Figure~\ref{fig:ratio}. 

From the figure, we observe that the enhancement in extracted energy enabled by our new protocol is a universal feature across all values of $\kappa>0$. This enhancement is not due to initializing the system in an excited state for the new protocol, as~opposed to the ground-state initialization in the original protocol. In~fact, for~both protocols, Bob’s local state after Alice’s measurement is ``excited'' in terms of the corresponding local effective Hamiltonian. The~true source of the enhancement lies in the fact that $\sin^2(\theta)$ in Equation~\eqref{extract} is greater than $\sin^2(\phi - \theta)$ in Equation~\eqref{extractold}. This arises because, in~the new protocol, Bob’s departure states coincide---i.e., they are independent of Alice’s measurement outcome. The~coincidence of the departure states leads to a larger distance from the destination states, thereby allowing more energy to be extracted. This coincidence arises from the absence of initial entanglement, such that Alice's measurement does not affect Bob's local states. We thus demonstrate that relaxing the entanglement constraint can enhance the performance of the QET protocol in terms of energy~extraction.

As a side remark, we noticed a similar study~\cite{haque2024aspects} that explores QET using the same Hamiltonian, Equation~\eqref{flipflop}, as~ours. In~their analysis, the~entangled ground state \((|01\rangle - |10\rangle)/\sqrt{2}\) was chosen as the system’s initial state. However, as~has been discussed, the~post-measurement state derived from this ground state does not exhibit the SLP property, making QET protocols unnecessary for Bob’s energy extraction. Instead, the~excited state \(|00\rangle\) must be used to apply QET. Furthermore, since their approach does not incorporate our developed concept of local effective Hamiltonians, they rely on optimization methods to determine Bob’s unitary operations for optimal energy extraction. This further highlights the advantages of our~results.

In this section, we presented a particular QET protocol that relaxes all three constraints. In~the Appendix \ref{appb} 
we further show how to construct a family of such protocols that also relax these constraints. This demonstrates that the phenomenon of locally inaccessible energy is widespread in the entire quantum Hilbert space, implying a broader necessity and applicability of QET~protocols.

\section{Experimental Verification of QET Protocol on Quantum~Hardware}
\begin{figure}[t]
    \includegraphics[width=0.6\linewidth]{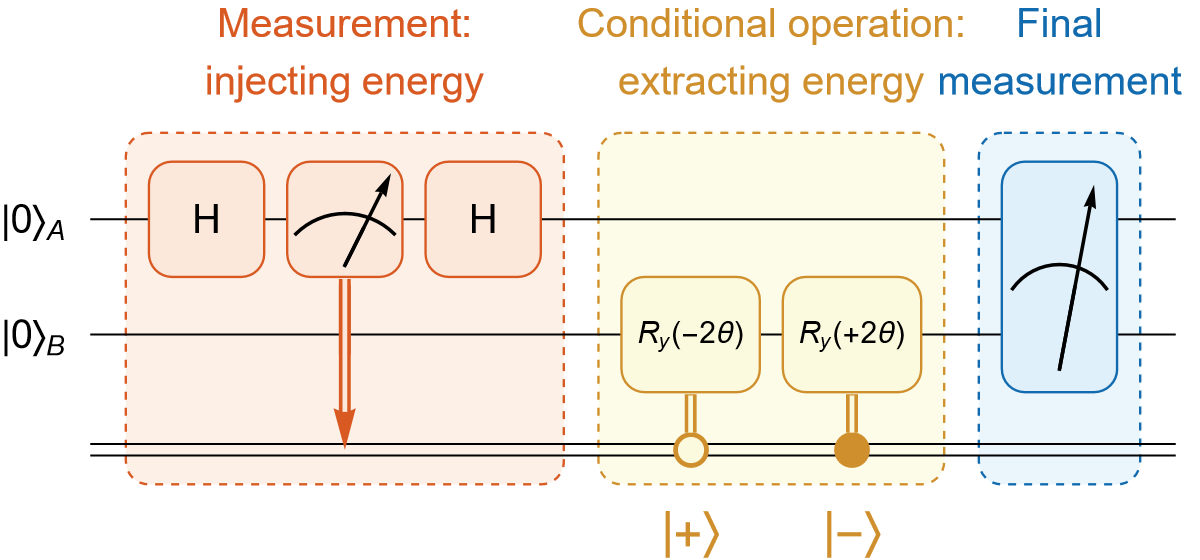}
    \caption{{The quantum} circuit implemented on IBM quantum hardware to verify the feasibility of our proposed QET protocol. Since the system begins in the first excited state of the Hamiltonian in Equation~\eqref{flipflop}, $|0\rangle \otimes |0\rangle$, no additional state preparation is required. Alice initiates the protocol by performing a measurement, which injects energy into the system. After~receiving Alice’s measurement outcome, Bob applies a conditional operation to extract energy from the system. A~final measurement is then performed to verify the energy distribution within the system.}
    \label{fig:circuit1}
\end{figure}

To validate the feasibility of our proposed QET protocol and demonstrate that the three traditional constraints are unnecessary, we implement the protocol on IBM quantum hardware. The~quantum circuit used for this implementation is shown in Figure~\ref{fig:circuit1}, which includes mid-circuit measurements, often referred to as dynamic~circuits.

Recently, IBM hardware has removed its support for dynamic circuits. To~address this limitation, we adopt the approach introduced in~\cite{ikeda2023demonstration}, replacing Bob's conditional operations with two-qubit controlled operations while delaying Alice's measurement until after these controlled operations. It can be shown that the two approaches are equivalent in the sense that they produce the same final density~matrix. 

To show this, suppose that Bob's operation is $U(\pm)$, conditional on Alice's measurement outcome $\pm$. After~Alice's measurement and Bob's operation, the~state becomes
\begingroup
\makeatletter\def\f@size{9.5}\check@mathfonts
\def\maketag@@@#1{\hbox{\m@th\normalsize\normalfont#1}}
\begin{equation}\label{eqn1}
    \rho \rightarrow \Big(|+\rangle\langle+| \otimes U(+)\Big) \rho \Big(|+\rangle\langle+| \otimes U^\dagger(+)\Big) 
    + \Big(|-\rangle\langle-| \otimes U(-)\Big) \rho \Big(|-\rangle\langle-| \otimes U^\dagger(-)\Big).
\end{equation}
\endgroup

Alternatively, in~the modified quantum circuit, the~two-qubit controlled unitary operations are applied first. The~state is first transformed as
\begin{equation}\label{eqn2}
    \rho \rightarrow \Big(|+\rangle\langle+| \otimes U(+) + |-\rangle\langle-| \otimes U(-)\Big) 
    \rho \Big(|+\rangle\langle+| \otimes U^\dagger(+) + |-\rangle\langle-| \otimes U^\dagger(-)\Big).
\end{equation}
After this transformation, Alice performs her delayed measurement. This measurement projects the state in Equation~\eqref{eqn2} into the same final state as Equation~\eqref{eqn1}, showing the equivalence of the two~approaches. 

In our case, the~original quantum circuit for our protocol is shown in Figure~\ref{fig:circuit1}, where the initial state is $|0\rangle_A \otimes |0\rangle_B$. The~red block represents Alice's measurement in the basis $\{|+\rangle, |-\rangle\}$. The~yellow block represents Bob's conditional operation based on Alice's classical information. The~blue block corresponds to the final measurement used to check the~energy. 

\begin{figure}[t]
    \includegraphics[width=0.6\linewidth]{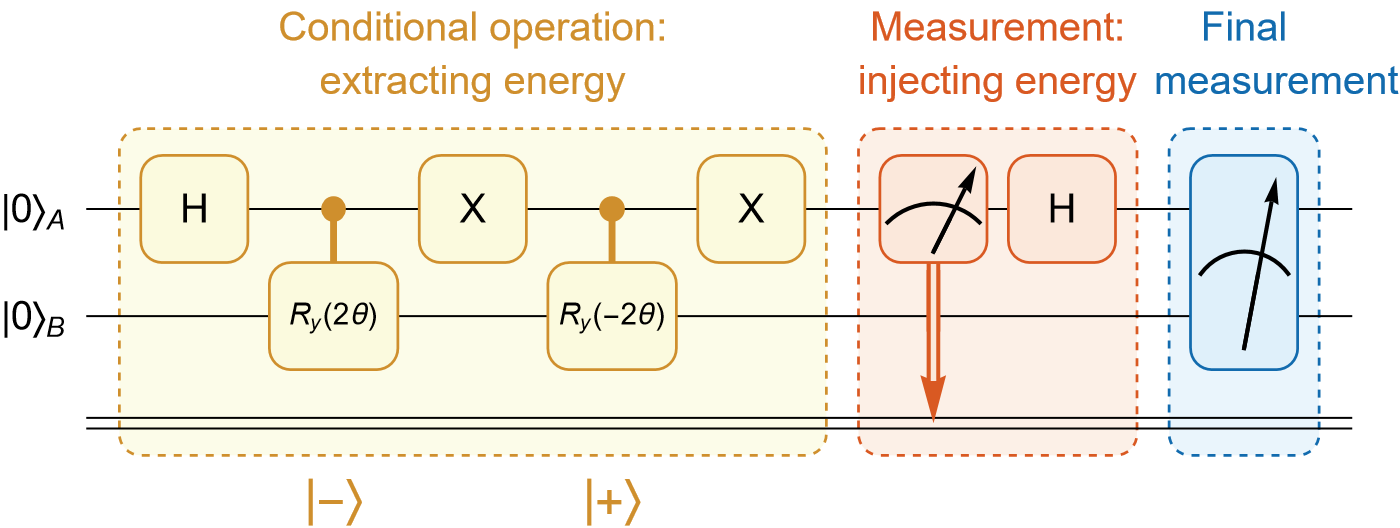}
    \caption{{To address} the current lack of support for dynamic circuits by IBM quantum hardware, we use an alternative but equivalent circuit. In~this approach, Alice’s measurement is postponed until after Bob’s conditional operations, and~Bob’s conditional operations are replaced with two-qubit controlled operations. It can be shown that the two circuits are equivalent, as~they yield identical resulting density~matrices.}
    \label{fig:circuit2}
\end{figure}

Instead, we replace the original circuit with the modified version shown in Figure~\ref{fig:circuit2}. In~this circuit, the~yellow block is first implemented, which corresponds to Bob's controlled operation based on Alice's quantum state in her computational basis. The~red block now represents Alice's delayed measurement. The~two circuits are equivalent, as they yield the same final density matrix and identical measurement outcomes for any~observables.

For the choice of parameters, without~loss of generality, we choose $h=1$. Then we choose $\kappa=1.5$, which satisfies the relation $\kappa>h$. By~substituting these into Equation~\eqref{extract}, we have the analytical value of the extracted energy as $E_\text{extract}=0.8028$.

\unskip

Next, for~the number of shots, we use 20,000 shots for each observable measurement, as~can be justified by applying Chebyshev's inequality:
\begin{equation}
    P\big(|\overline{\mathcal{O}} - \langle \mathcal{O} \rangle| \geq \varepsilon\big) \leq \frac{\text{Var}(\mathcal{O})}{N\varepsilon^2} \leq \delta,
\end{equation}  
where $\overline{\mathcal{O}}$ is the statistical mean obtained from the measurements of an arbitrary observable $\mathcal{O}$, and~$\langle \mathcal{O} \rangle$ is its theoretical expectation value. In~our case, $\mathcal{O}$ is always a Pauli word (product of individual Pauli operators), so $\text{Var}(\mathcal{O})$ is upper-bounded by 1. This gives the relation $\delta \geq 1 / (N\varepsilon^2)$.  

For \( N = \text{20,000} \), if~we set \( \varepsilon = 0.02 \), we find \( \delta = 1/8 \). This implies that there is at most a \( 1/8 \) probability that our measurement outcome deviates from the true value by more than \( 0.02 \). This level of accuracy is sufficient for our experimental~purposes.  

For the backends, we choose three available IBM hardware backends: \texttt{ibm\_brussels}, \texttt{ibm\_kyiv}, and~\texttt{ibm\_torino}. Detailed specifications for these backends, along with their qubit distribution maps, are provided in Appendix \ref{appc}. Additionally, we execute the circuits using the \texttt{AerSimulator}, a~classical simulator capable of running quantum circuits locally on personal computers. When no noise is introduced, the~\texttt{AerSimulator} is expected to produce exact results, serving as a benchmark for the design of the quantum~circuits.

The measurement results are listed in Table~\ref{tab:result1}. In~the left table, we report the measurements of the three operators $\mathbb{I}_A \otimes Z_B$, $X_A \otimes X_B$, and~$Y_A \otimes Y_B$ prior to Bob applying his conditional operations. In~the right table, these operators are measured after Bob has applied his conditional operations. The~energy that we are interested in is given by
\begin{equation}
E_\text{Bob} = -h\langle \mathbb{I}_A \otimes Z_B \rangle + \kappa\langle X_A \otimes X_B \rangle + \kappa\langle Y_A \otimes Y_B \rangle.
\end{equation}
The term $-h\langle Z_A \otimes \mathbb{I}_B \rangle$ is omitted from our calculation, as~it remains unaffected by Bob's conditional operations. $E_\text{Bob}$ is compared before and after Bob's conditional operations. If~the energy in the right table is smaller than that in the left table, it indicates that Bob's conditional operation has successfully extracted energy from the~system. 

\begin{table}[ht]
\centering
\begin{footnotesize}

\begin{minipage}{0.5\textwidth}
    \centering
        \begin{tabular}{lcc}
        \hline\hline\\[-0.8em]
        Backend & Operator & Result \\[0.2em]
        \hline\\[-0.8em]
        Analytical &  \multirow{5}{*}{$\langle \mathbb{I}_A\otimes Z_B\rangle$}   &   1.0\\
        AerSimulator &    &   1.0\\
        ibm\_brussels &     &  $1.0081\pm0.0072$ \\
        ibm\_kyiv &    &  $1.0035\pm0.0034$ \\
        ibm\_torino &     & $1.1015\pm0.0195$  \\[0.2em]
        \hline\\[-0.8em]
        Analytical &  \multirow{5}{*}{$\langle X_A\otimes X_B\rangle$}   &   0.0\\
        AerSimulator &    &   0.0\\
        ibm\_brussels &     &  $0.0029\pm0.0073$ \\
        ibm\_kyiv &   & $0.0121\pm0.0064$  \\
        ibm\_torino &     &  $-0.0062\pm0.0106$ \\[0.2em]
        \hline\\[-0.8em]
        Analytical &  \multirow{5}{*}{$\langle Y_A\otimes Y_B\rangle$}  &   0.0\\
        AerSimulator &    &   0.0\\
        ibm\_brussels &    &  $0.0185\pm0.0083$ \\
        ibm\_kyiv &   &  $0.0001\pm0.0060$ \\
        ibm\_torino &     &  $0.0153\pm0.0095$ \\[0.2em]
        \hline\\[-0.8em]
        Analytical &  \multirow{5}{*}{$\langle E_\text{Bob}\rangle$}  &   $-1.0$\\
        AerSimulator &    &   $-1.0$\\
        ibm\_brussels &    &  $-0.9760\pm0.0161$ \\
        ibm\_kyiv &   &  $-0.9852\pm0.0152$ \\
        ibm\_torino &     &  $-1.0878\pm0.0107$ \\[0.2em]
        \hline\hline
        \end{tabular}
\end{minipage}%
\hfill
\begin{minipage}{0.5\textwidth}
    \centering
    \begin{tabular}{lcc}
        \hline\hline\\[-0.8em]
        Backend & Operator &  Result \\[0.2em]
        \hline\\[-0.8em]
        Analytical &  \multirow{5}{*}{$\langle\mathbb{I}_A\otimes Z_B\rangle$}   &   0.5547\\
        AerSimulator &   &    $0.5547$\\
        ibm\_brussels &    &  $0.5461\pm0.0068$ \\
        ibm\_kyiv &   &  $0.5217\pm0.0060$ \\
        ibm\_torino &     &  $0.6515\pm0.6515$ \\[0.2em]
        \hline\\[-0.8em]
        Analytical &  \multirow{5}{*}{$\langle X_A\otimes X_B\rangle$}  &   $-0.8320$\\
        AerSimulator &   &   $-0.8320$\\
        ibm\_brussels &     &  $-0.6361\pm0.0085$ \\
        ibm\_kyiv &    &  $-0.7697\pm0.0043$ \\
        ibm\_torino &    & $-0.8495\pm-0.8495$  \\[0.2em]
        \hline\\[-0.8em]
        Analytical &  \multirow{5}{*}{$\langle Y_A\otimes Y_B\rangle$}   &   0.0\\
        AerSimulator &   &   0.0\\
        ibm\_brussels &     &  $0.0069\pm0.0072$ \\
        ibm\_kyiv &    & $-0.0129\pm0.0074$   \\
        ibm\_torino &     &  $0.0105\pm0.0105$ \\[0.2em]
        \hline\\[-0.8em]
        Analytical &  \multirow{5}{*}{$\langle E_\text{Bob} \rangle $}   &   $-1.8028$\\
        AerSimulator &   &   $-1.8028$\\
        ibm\_brussels &     &  $-1.4899\pm0.0168$ \\
        ibm\_kyiv &    & $-1.6956\pm0.0115$   \\
        ibm\_torino &     &  $-1.9100\pm0.0364$ \\[0.2em]
        \hline\hline
    \end{tabular}
\end{minipage}
\end{footnotesize}
\caption{The measurement results for the QET protocol are presented in two tables. The left table shows the results obtained before Bob's conditional operations, which corresponds to removing the yellow block in Fig.~\ref{fig:circuit1} while retaining the rest of the circuit. The right table shows the results after Bob's conditional operations, corresponding exactly to the circuit shown in Fig.~\ref{fig:circuit2}. Three operators are measured: $\mathbb{I}_A \otimes Z_B$, $X_A \otimes X_B$, and $Y_A \otimes Y_B$, yielding the final value of $E_\text{Bob}$, representing Bob's energy. The difference between the results in the two tables signifies the energy extracted by Bob's conditional operations, as are further given in Table~\ref{tab:result2}.
}
\label{tab:result1}
\end{table}

The standard deviations in these measurements arise only from statistical fluctuations and do not account for systematic errors introduced by the near-term intermediate-scale quantum (NISQ) nature of IBM hardware. To~minimize systematic errors, we use the measurement error mitigation technique~\cite{qiskit2025}, implemented as a built-in function within IBM's \texttt{qiskit} package, by~setting the resilience level to 1. This approach effectively reduces errors occurring during the measurement process of the quantum circuit. These corrections are directly incorporated into the expectation values of each measurement outcome, while the standard deviations only reflect statistical~uncertainty.

\begin{table}[ht]
    \centering
    \begin{footnotesize}
    \begin{tabular}{cccccc}
    \hline\hline\\[-0.8em]
        Backends    & Analytical  & AerSimulator  & ibm\_brussels  & ibm\_kyiv  & ibm\_torino  \\[0.2em]
        \hline\\[-0.8em]
        Extracted energy   &  0.8028 &  0.8028 &  $0.5139\pm0.0233$ & $0.7104\pm0.0191$  & $0.8222\pm0.0379$  \\[0.2em]
    \hline\hline
    \end{tabular}
    \end{footnotesize}
    \caption{The energy extracted by Bob is presented for the three IBM quantum hardware backends, along with results from the AerSimulator and the analytical solution. All three quantum hardware backends successfully extract energy from the system, thereby breaking the restriction of strong local passivity and confirming the feasibility of our proposed QET protocol.
}
    \label{tab:result2}
\end{table}

By comparing the results in the two panels, the~extracted energy can be calculated and is presented in Table~\ref{tab:result2}. The~standard deviation for the extracted energy is determined using the error propagation formula. Specifically, since $E_\text{extract} = E_\text{Bob}^\text{left} - E_\text{Bob}^\text{right}$, the~standard deviation is given by $\Delta E_\text{extract} = \sqrt{(\Delta E_\text{Bob}^\text{left})^2 + (\Delta E_\text{Bob}^\text{right})^2}$. Notably, the~\texttt{ibm\_torino} backend produced results closer to the analytical solution compared to the other two backends. This improvement is attributed to the newer Heron processor used by \texttt{ibm\_torino}, whereas the other two backends use Eagle processors. Further details about these processors are provided in Appendix~\ref{appc}.

The results demonstrate that all three quantum hardware backends successfully provide positive extracted energy. This confirms the experimental feasibility of our newly proposed QET protocol and further establishes the fact that the three previously considered constraints for QET are~unnecessary.

Regarding our experimental setup in Figure~\ref{fig:circuit2}, one might initially wonder why the Hamiltonian term appears to be missing from the circuit and, consequently, how energy can be extracted from the quantum system. We clarify that our implementation uses quantum hardware to simulate the QET process. The~effect of a Hamiltonian \( H \) in a physical process is fully captured by its time evolution operator, \( U \equiv e^{-iHt} \), which can be efficiently implemented using quantum circuits to simulate the evolution of qubit states under their~interaction.  

Importantly, the~QET protocol assumes nonrelativistic conditions, allowing signals to propagate much faster than the system’s natural evolution. Furthermore, all steps in a QET protocol---including Alice’s initial measurement, Bob’s conditional operations, and~final energy measurement---can, in~principle, occur on timescales much shorter than that of the Hamiltonian \( H \). As~a result, the~system's time evolution during these steps is approximately the identity operator, \( U \approx \mathbb{I} \), making the Hamiltonian’s evolutional effect negligible in our circuits. Nevertheless, this does not undermine the validity of our quantum simulation, which successfully demonstrates local energy extraction enabled by QET, as~can be observed in Table~\ref{tab:result1}.

\section{Summary}

Quantum state teleportation (QST) is a well-established protocol for transferring quantum states between distant locations. However, the~energy required to reconstruct the quantum state is provided by the receiver (Bob), rather than being transmitted directly from the sender (Alice). As~a result, QST is not capable of teleporting energy. To~address this, the~quantum energy teleportation (QET) protocol was introduced. However, unlike QST, the~scenarios where QET is needed impose stricter constraints. These include the need for the initial state shared by Alice and Bob to be entangled and to correspond to the ground state of an interacting Hamiltonian. Additionally, the~observable used by Alice for her measurement must commute with the interaction term in the Hamiltonian. These constraints are overly restrictive and limit the broader applicability of the~protocol.

A key understanding of the QET protocol is that no energy is actually teleported from Alice to Bob within the protocol. Instead, Alice's message acts as a key that enables Bob to access and extract his local energy, which was previously inaccessible due to strong local passivity (SLP). In~light of this insight, the~previously considered ``reasonable'' constraints for QET appear unnecessary. Nevertheless, these strict conditions are still widely regarded as essential in recent discussions of QET~protocols.

In this work, we demonstrate that SLP can arise beyond these conventional constraints, establishing the necessity of QET in a wider range of scenarios for local energy extraction. Specifically, Alice and Bob begin with a product state, which corresponds to an excited eigenstate of the Hamiltonian rather than its ground state. Furthermore, Alice's measurement does not commute with the interaction term of the Hamiltonian. Despite this, we demonstrate that the system maintains SLP both before and after Alice's measurement. This ensures that Bob cannot locally extract energy from the system in a consistent way. Instead, Alice communicates her measurement outcome to Bob, enabling him to perform conditional operations. These conditional operations allow Bob to extract energy, effectively performing the QET protocol without being limited by those~constraints.

Additionally, the~amount of energy that can be extracted using our new protocol is amplified to be 7.2 times higher than that of the original protocol, highlighting a significant advantage of the new protocol in retrieving otherwise unavailable~energy.

Furthermore, we implement our new QET protocol on three available IBM quantum hardware backends. The~results from all three backends demonstrate positive extracted energy for Bob, validating the feasibility of our protocol in experimental settings and confirming that the three previously considered constraints are not necessary. These successful implementations confirm the effectiveness of our new protocol, providing deeper insights into quantum energy teleportation and enabling its broader applications by removing unnecessary~constraints.

In conclusion, our results suggest that the phenomenon of locked energy due to SLP extends beyond the systems initially anticipated by researchers. These locked energies, which cannot be locally extracted, limit our ability to leverage them for practical applications. However, our results show that these inaccessible energies can be unlocked and extracted using QET protocols with classical communications, making them available for future~use. 

The advantages of our results~include the following:
\begin{itemize}
    \item First, it clarifies the key connection between QET and SLP.
    \item Second, by~relaxing previous constraints, it expands the broader applicability of QET.
    \item Third, the~introduction of the local effective Hamiltonian brings technical merits.
    \item Fourth, our new QET protocol amplifies the amount of extracted energy---7.2 times more than that in the original protocol---thereby enhancing the power of QET.
\end{itemize}

Potential future directions include extending the current protocols to many-body systems, which would further broaden the applicability of our proposed concept of the ``local effective Hamiltonian'' in more complex quantum settings. Another promising direction is to explore practical applications of the extracted energy in various quantum systems---for example, in~quantum chemistry, where such energy extraction could enable novel approaches to modeling chemical reactions. Pursuing these avenues may yield significant advances in both fundamental understanding and technological~applications.


\vspace{6pt}

\acknowledgments{We would like to thank Peter W.~Milonni and the late Joseph H.~Eberly for valuable discussions. We acknowledge two funding sources from the U.S. Department of Energy (DOE). The first is provided by the Office of Science through the Quantum Science Center (QSC), a National Quantum Information Science Research Center; the second is provided by the Office of Basic Energy Sciences (BES) under Award DE-SC0025620.}


\appendix
\section[\appendixname~\thesection]{Analytical Form of the Matrix \boldmath$M(\rho, H)$}\label{appa}
This is a repetition of the main result given in~\cite{alhambra2019fundamental}. Given a Hamiltonian \( H_{AB} \) and a two-party state \( \rho_{AB} \), we define the energy difference under a local completely positive trace-preserving (CPTP) map acting on subsystem \( B \) as
\begin{equation}
    \Delta E_{A(B)} = \min_{\mathcal{G}_B} \text{Tr}[H_{AB} (\mathcal{I}_A \otimes \mathcal{G}_B)\rho_{AB}] - \text{Tr}(H_{AB} \rho_{AB}).    
\end{equation}

It is easy to verify that the~pair \( \{\rho_{AB}, H_{AB}\} \) is strong local passive (SLP) with respect to subsystem \( B \) if and only if this energy difference vanishes:
\begin{equation}
    \Delta E_{A(B)} = 0.    
\end{equation}

Now, consider an auxiliary Hilbert space \( B' \) with the same dimension as \( B \). Define the Hermitian operator \( C_{BB'} \in \mathcal{H}_B \otimes \mathcal{H}_{B'} \) as
\begin{equation}
    C_{BB'} = \text{Tr}_A\left[\rho_{AB}^{\Gamma_B} H_{AB'}\right],    
\end{equation}
where \( \rho_{AB}^{\Gamma_B} \) denotes the partial transpose of \( \rho_{AB} \) on subsystem \( B \). The~Choi--Jamiołkowski operator for the identity channel is given by
\begin{equation}
    \mathcal{J}^\mathcal{I}_{BB'} \equiv 2 |\Phi^+\rangle_{BB'}\langle\Phi^+|_{BB'} = \big(|0\rangle_B |0\rangle_{B'} + |1\rangle_B |1\rangle_{B'}\big)\big(\langle0|_B \langle0|_{B'} + \langle1|_B \langle1|_{B'}\big).    
\end{equation}

Finally, it is concluded that \( \Delta E_{A(B)} = 0 \) if and only if the Hermitian operator
\begin{equation}
    M(\rho_{AB}, H_{AB}) \equiv C_{BB'} - \text{Tr}_{B'}\left[\mathcal{J}^\mathcal{I}_{BB'} C_{BB'}\right] \otimes \mathbb{I}_{B'}    
\end{equation}
is positive semidefinite, i.e.,
\begin{equation}
    M(\rho_{AB}, H_{AB}) \geq 0.    
\end{equation}

Or equivalently, all four of its eigenvalues are nonnegative. The~proof of this result can be found in~\cite{alhambra2019fundamental}.

\section[\appendixname~\thesection]{Construction of a Class of QET Protocols}\label{appb}
While the main text presents one specific QET protocol that goes beyond the three discussed constraints, here we show that a broader class of such protocols can be constructed. However, we emphasize that this construction is not unique and does not cover all possible constructions with the same~desire.

We start from the spectrum of a two-qubit Hamiltonian, with~its four eigenvalues and eigenvectors given~by
\begin{table}[h]
    \centering
    \begin{tabular}{c|c}
    \hline\hline\\[-0.5em]
    Eigenvalues & Eigenvectors\\[0.5em]
    \hline\\[-0.5em]
    $\mathcal{E}_4$& $|v_4\rangle=|\psi^\perp\rangle\otimes|\phi^\perp\rangle$ \\[0.5em]
    \hline\\[-0.5em]
       $\mathcal{E}_3$  & $|v_3\rangle=|1\rangle\otimes|\phi\rangle$ \\[0.5em]
    \hline\\[-0.5em]
       $\mathcal{E}_2$  & $|v_2\rangle=|0\rangle\otimes|\phi\rangle$ \\[0.5em]
     \hline\\[-0.5em]
    $\mathcal{E}_1$& $|v_1\rangle=|\psi\rangle\otimes|\phi^\perp\rangle$ \\[0.5em]
    \hline\hline
    \end{tabular}
\end{table}

\noindent where $\mathcal{E}_1<\mathcal{E}_2<\mathcal{E}_3<\mathcal{E}_4$ is assumed. Here, the~single-qubit states are defined as
\begin{equation}
    \begin{split}
        |\phi\rangle=&\cos(\alpha)|0\rangle+\sin(\alpha)|1\rangle\\
        |\phi^\perp\rangle=&\sin(\alpha)|0\rangle-\cos(\alpha)|1\rangle\\
        |\psi\rangle=&\cos(\beta)|+\rangle+\sin(\beta)|-\rangle\\
        |\psi^\perp\rangle=&\sin(\beta)|+\rangle-\cos(\beta)|-\rangle\\
    \end{split}
\end{equation}
Therefore, the~total Hamiltonian is
\begin{equation}\label{habappendix}
    H_{AB}=\mathcal{E}_1|v_1\rangle\langle v_1|+\mathcal{E}_2|v_2\rangle\langle v_2|+\mathcal{E}_3|v_3\rangle\langle v_3|+\mathcal{E}_4|v_4\rangle\langle v_4|.
\end{equation}

We set the initial state to be $|v_2\rangle$, which is an excited state with no entanglement. And~we let Alice measure her qubit in the Pauli-$X$ basis. This leads to Bob's post-measurement states being the following, each with 50\% probability:
\begin{equation}\label{postappendix}
    \begin{split}
        |b^+\rangle=&|+\rangle\otimes|\phi\rangle,\\
        |b^-\rangle=&|-\rangle\otimes|\phi\rangle.\\
    \end{split}
\end{equation}

At this stage, for~simplicity, we choose the energies $\mathcal{E}_4=-\mathcal{E}_1=\mathcal{E}$ and $\mathcal{E}_3=-\mathcal{E}_2=\mathcal{F}$. The~readers may make their own choices, but~the analyses can be more~complex.

Under the above choice of eigenvalues, Bob's local effective Hamiltonians are given by
\begin{equation}
    \begin{split}
        H_\text{eff}^+=&\langle+|H_{AB}|+\rangle=[\sin^2(\beta)-\cos^2(\beta)]\mathcal{E}|\phi^\perp\rangle\langle\phi^\perp|,\\
        H_\text{eff}^-=&\langle-|H_{AB}|-\rangle=[\cos^2(\beta)-\sin^2(\beta)]\mathcal{E}|\phi^\perp\rangle\langle\phi^\perp|=-H_\text{eff}^+.
    \end{split}
\end{equation}

Because of the equality $\langle\phi|\phi^\perp\rangle=0$, it can be seen that Bob's local state $|\phi\rangle$ is an eigenstate of both effective Hamiltonians $H_\text{eff}^\pm$: it serves as the ground state in one case and the excited state in the other, depending on the choice of $\beta$. As~a result, any local operation performed by Bob cannot extract energy on average: if energy is extracted in one case, an~equal amount is injected in the other due to symmetry. This implies that the post-measurement state satisfies the SLP condition, and~Bob cannot locally extract energy after Alice's~measurement.

Furthermore, it is evident that classical communication is required to enable Bob’s local energy extraction. Upon~receiving Alice’s message, Bob learns whether his local state is the ground state or the excited state. If~it is the ground state, he performs no operation. If~it is the excited state, he applies a $\pi$ rotation to bring the state to the ground state $|\phi^\perp\rangle$, thereby extracting~energy.

In fact, one can numerically show that for any choice of $\alpha$ and $\beta$ in the Hamiltonian $H_{AB}$, and~with the help of the method described in~\cite{alhambra2019fundamental}, the~initial state $|v_2\rangle\langle v_2|$ is an SLP state with respect to $H_{AB}$, and~that the post-measurement density matrix obtained from Equation~\eqref{postappendix} also satisfies the SLP condition for $H_{AB}$. This construction therefore yields a two-parameter family of Hamiltonians that demonstrate the SLP property both for an excited initial state and for its corresponding post-measurement state, with~entanglement entirely absent throughout the process. Also, in~general, the~Pauli-$X$ basis that Alice uses for her measurement does not necessarily commute with the interaction Hamiltonian within Equation~\eqref{habappendix}. This demonstrates that all three constraints are relaxed for this general construction. It therefore highlights the necessity of QET for enabling local energy extraction in broader scenarios and~applications.

\section[\appendixname~\thesection]{Details of the Quantum Hardware}\label{appc}

\begin{figure}[h]
    \includegraphics[width=0.7\linewidth]{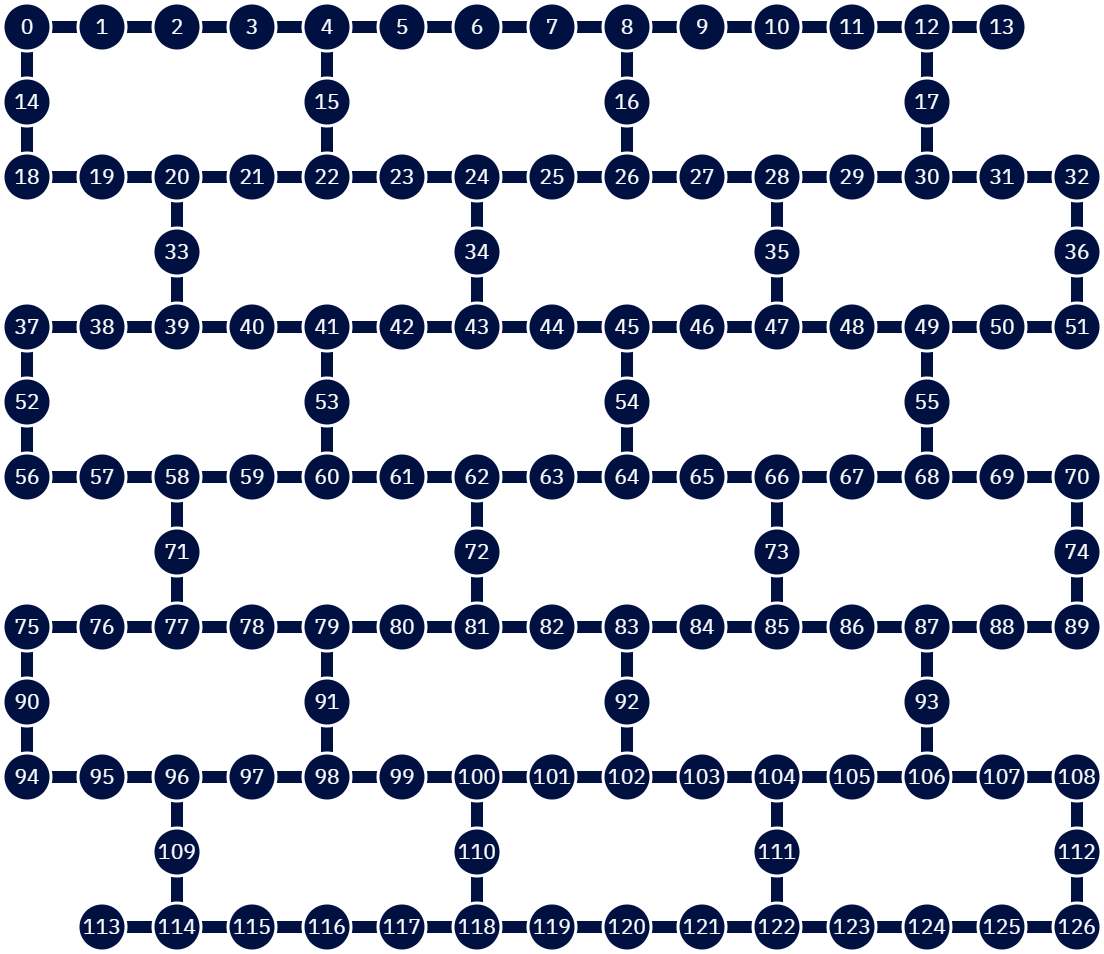}
    \caption{The qubit distribution map for the backends of \texttt{ibm\_brussels} and \texttt{ibm\_kyiv} is demonstrated. Qubit 0 is used for Alice's qubit, and~qubit 1 is used for Bob's~qubit.}
    \label{fig:brussel}
\end{figure}

\begin{figure}[h]
    \includegraphics[width=0.7\linewidth]{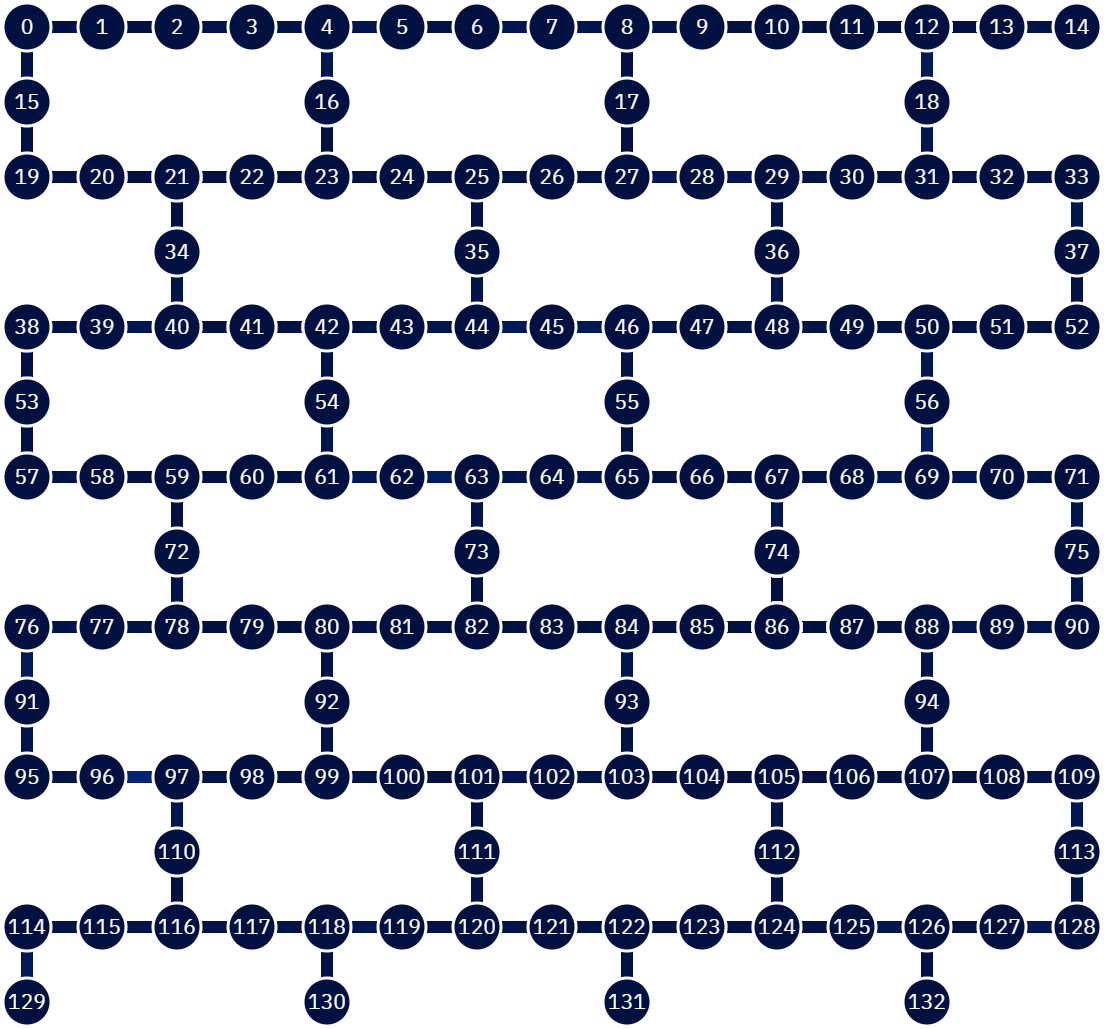}
    \caption{The qubit distribution map for the backend of \texttt{ibm\_torino} is demonstrated. Qubit 0 is used for Alice's qubit, and~qubit 1 is used for Bob's~qubit.}
    \label{fig:torino}
\end{figure}

The details of the three IBM quantum hardware backends are provided in this section. The~qubit distribution maps for \texttt{ibm\_brussels} and \texttt{ibm\_kyiv} are shown in Figure~\ref{fig:brussel}, while the map for \texttt{ibm\_torino} is presented in Figure~\ref{fig:torino}. In~all three quantum circuits, qubit 0 is used for Alice's qubit, and~qubit 1 is used for Bob's~qubit.

Table~\ref{tab:machine} provides detailed information about the three IBM quantum hardware backends, including the properties of the qubits used and the properties of the single-qubit and two-qubit gates employed in the experiments. This information was retrieved from~\cite{qiskit2025} on 13 January 2025, the~day when the quantum circuits were executed on these~backends.


\unskip

\unskip

\begin{table}[h]
    \centering
    \begin{small}
    \begin{tabular}{lccc}
    \hline\hline\\[-0.8em]
    & \multicolumn{3}{c}{\bf Backend}\\[0.2em]
    \hline\\[-0.8em]
       & ibm\_brussels &  ibm\_kyiv &  ibm\_torino \\[0.2em]
    \hline\\[-0.8em]
    Processor type &  Eagle r3 &  Eagle r3 &  Heron r1  \\
    $N_\text{tot}$   & 127  &  127 & 133  \\
    Quantum volume     & 128  & 128  & 512  \\
    Shots & 20000  & 20000  &  20000  \\
    Readout length ($ns$) & 1500  & 1244  &  1560  \\
    Qubits used  &  2 & 2  & 2   \\
    ECR error  & 0.01274  & 0.00612  &  N/A  \\
    CZ error  &  N/A & N/A  &  0.00214   \\
    Gate time (ns) & 660  & 562  & 68   \\
    CLOPS &  220K &  30K & 210K   \\[0.2em]
    \hline\hline\\[-0.8em]
    & \multicolumn{3}{c}{\bf Alice's qubit}\\[0.2em]
    \hline\\[-0.8em]
    Qubit index & 0  & 0  & 0     \\
    $T_1$ ($\mu s$)  &  370.15 & 249.70  & 213.35    \\
    $T_2$ ($\mu s$)   & 105.38  & 307.57  &  319.17   \\
    Frequency(GHz)  & 4.898  & 4.656  & N/A    \\
    Anharmonicity (GHz)  &  $-0.30830$ &  $-0.31106$ &  N/A   \\
    Pauli-X error  &  $1.859\times10^{-4}$ & $3.073\times10^{-4}$  & $3.792\times10^{-4}$    \\
    Readout assignment error& 0.0414  & 0.0035  &  0.1248  \\[0.2em]
    \hline\hline\\[-0.8em]
    & \multicolumn{3}{c}{\bf Bob's qubit}\\[0.2em]
    \hline\\[-0.8em]
    Qubit index & 1  & 1 &  1    \\
    $T_1$ ($\mu s$)  & 233.54  & 412.87  & 291.45   \\
    $T_2$ ($\mu s$)   & 250.55  &  200.94 &  194.08   \\
    Frequency(GHz)  &  4.839 & 4.535  &  N/A   \\
    Anharmonicity (GHz)  & N/A  &  $-0.31303$ &  N/A   \\
    Pauli-X error  & $2.487\times10^{-4}$  & $1.035\times10^{-4}$  & $1.398\times10^{-4}$   \\
    Readout assignment error&  0.0177 & 0.0028  &  0.0424   \\[0.2em]
    \hline\hline
    \end{tabular}
    \end{small}
    \caption{The machine properties of the IBM quantum hardware and the parameters used in our experiments are summarized. ``Shots'' refers to the number of iterations performed for sampling. ``CLOPS'' refers to the number of circuit layer operations per second, representing how many layers of a circuit the quantum processing unit can execute per unit time. ``ECR error'' refers to the error rate for the two-qubit ECR gate, employed by the Eagle-type processors, while ``CZ error'' refers to the error rate for the two-qubit CZ gate, employed by the Heron-type processors. Note that the frequency and anharmonicity information of the qubits for the Heron-type processors is unavailable from the IBM workload source and is therefore not included here.
}
    \label{tab:machine}
\end{table}

\end{document}